\documentclass[12pt]{article}
\usepackage{amsmath}
\usepackage{times}
\usepackage{graphicx}
\usepackage{adjustbox}
\usepackage{color}
\usepackage[dvipsnames]{xcolor}
\usepackage{multirow}
\usepackage[normalem]{ulem}
\usepackage{siunitx}
\usepackage[authoryear]{natbib}
\usepackage{rotating}
\usepackage{bbm}
\usepackage{latexsym}
\usepackage{subfig}

\newcommand{\beginsupplement}{%
        \setcounter{table}{0}
        \renewcommand{\thetable}{S\arabic{table}}%
        \setcounter{figure}{0}
        \renewcommand{\thefigure}{S\arabic{figure}}%
     }

\newcommand{\captionfonts}{\normalsize}

\makeatletter  
\long\def\@makecaption#1#2{%
  \vskip\abovecaptionskip
  \sbox\@tempboxa{{\captionfonts #1: #2}}%
  \ifdim \wd\@tempboxa >\hsize
    {\captionfonts #1: #2\par}
  \else
    \hbox to\hsize{\hfil\box\@tempboxa\hfil}%
  \fi
  \vskip\belowcaptionskip}
\makeatother

\newsavebox{\measurebox}

\begin{document}
\hspace{13.9cm}1

\ \vspace{20mm}\\

{\LARGE NetPyNE implementation and rescaling of the Potjans-Diesmanncortical microcircuit model}

\ \\
{\bf Cecilia Romaro$^{1}$, Fernando Araujo Najman$^{2}$, William W Lytton$^{3}$, Antonio C Roque$^{1}$, Salvador Dura-Bernal$^{3,4}$}\\
{$^{1}$Department of Physics, School of Philosophy, Sciences and Letters of Ribeir\~{a}o Preto, University of S\~{a}o Paulo, Ribeir\~{a}o Preto, S\~{a}o Paulo, Brazil}\\
{$^{2}$Institute of Mathematics and Statistics, University of S\~{a}o Paulo, S\~{a}o Paulo, Brazil}\\
{$^{3}$Department of Physiology and Pharmacology, State University of New York, New York, NY, USA}\\
{$^{4}$ Nathan Kline Institute for Psychiatric Research, NY, USA}


\thispagestyle{empty}
\markboth{}{NC instructions}

\ \vspace{-0mm}\\

\begin{center} {\bf Abstract} \end{center}
The Potjans-Diesmann cortical microcircuit model is a widely used model originally implemented in NEST. Here, we re-implemented the model using  NetPyNE, a high-level Python interface to the NEURON simulator, and reproduced the findings of the original publication. We also implemented a method for rescaling the network size which preserves first and second order statistics, building on existing work on network theory. 
The new implementation enables using more detailed neuron models with multicompartment morphologies and multiple biophysically realistic channels. This opens the model to new research, including the study of dendritic processing, the influence of individual channel parameters, and generally multiscale interactions in the network. The rescaling method provides flexibility to increase or decrease the network size if required when running these more realistic simulations. Finally, NetPyNE facilitates modifying or extending the model using its declarative language; optimizing model parameters; running efficient large-scale parallelized simulations; and analyzing the model through built-in methods, including local field potential calculation and information flow measures.

\section{Introduction}

The Potjans-Diesmann cortical microcircuit (PDCM) model \citep{potjans2012cell} reproduces the cortical network under a 1 mm$^2$ surface area of early sensory cortex. The model generates spontaneous activity with layer-specific firing rates similar to those observed experimentally \citep{de2009spiking, sakata2009laminar, swadlow1989efferent}.

The PDCM model was the first one to reproduce the connectivity structure of the cortical layers with statistical fidelity to the biological data observed experimentally \citep{thomson2002synaptic, west2005layer}. This model is broadly used to study the emergence of macroscopic cortical patterns, such as layer specific oscillations \citep{van2015scalability, bos2016identifying} or effects on cortical functionality resulting from inter-layer or inter-columns communication \citep{cain2016computational, schwalger2017towards, schmidt2018multi}. Some examples of use of this model are the study of the influence of the microconnectome on the activity through the network layers \citep{schuecker2017fundamental}, modeling of spatial attention in the visual cortex \citep{wagatsuma2013spatial} and modeling the effects of inhibitory connections in contextual visual processing \citep{lee2017computational} and in different cortical microcircuitry regions  \citep{beul2015towards}. 

In this work, we converted the PDCM model from NEST to NetPyNE \citep{Dura19,lytton2016simulation} (www.netpyne.org). NetPyNE provides a high-level interface to the NEURON simulator \citep{carnevale2006neuron} that facilitates the development, parallel simulation and analysis of biological neuronal networks. 
NetPyNE provides a high-level declarative format that clearly separates the model parameters from the underlying implementation, making the PDCM model easier to understand, share and manipulate. NetPyNE enables efficient parallel simulation of the model with a single function call, and provides a wide array of built-in analysis functions to further explore the model. 

Our NetPyNE implementation enables employing more detailed cell models as alternatives to the original leaky integrate-and-fire (LIF) neurons. NetPyNE makes it possible to readily use PDCM model connection topology for more complex simulations by swapping in multicompartmental neuron models with arbitrarily detailed features: conductance-based channels, more complex synaptic models,  \citep{hines2004modeldb} and reaction-diffusion processes \citep{mcdougal2013reaction,ranjan2011channelpedia,Newt18}. This allows a new array of possible studies, such as investigating the interaction between network topology and dendritic morphology or channel-specific parameters \citep{Beza16c,Dura17b,Neym16b}.

More detailed simulations require considerable additional computational re--sources. To make these more detailed simulations computationally feasible, it may be necessary to reduce the number of neurons in the network. Given the increasing availability of supercomputing resources \citep{towns2014xsede,Siva13}, researchers may also wish to switch back and forth across different network sizes \citep{schwalger2017towards,schmidt2018multi,Beza16c}. However, rescaling the network to decrease or increase its size while maintaining its dynamical properties is a challenging process. For example, as we reduce the number of neurons we need to increase the number of connections or the synaptic weight to balance the external inputs. However, this can lead to an undesired spiking synchrony and regularity \citep{brunel2000dynamics}. To address this issue we implemented a rescaling method, adapted from the original model, to resize the number of network neurons, connections and external inputs as well as the synaptic weights, while keeping the matrix of connection probabilities and the proportions of cells per population fixed.

Our implementation is able to generate NEURON-based network models of different sizes with  layer-specific average firing rates, synchrony and irregularity features, similar to those in the original PDCM model \citep{potjans2012cell}. This will allow researchers to modify both the level of detail and size of the PDCM network to adapt it to their computational resources and research objectives.

\section{Methods}
\subsection{Original NEST PDCM model}
The network consists of around 80,000 leaky integrate-and-fire (LIF) \citep{lapicque1907recherches} neurons divided in eight cell populations representing excitatory and inhibitory neurons in cortical layers 2/3, 4, 5 and 6 (these populations will be referred to here by L2e, L2i, L4e, L4i, L5e, L5i, L6e and L6i). External input is provided by thalamic and cortico-cortical afferents.

The network, originally built in NEST \citep{gewaltig2007nest}, specifies fixed numbers of excitatory and inhibitory neurons per layer, the number and strength of connections between these neuronal populations and of external inputs to each cell population. These numbers are based on experimental data \citep{thomson2002synaptic, west2005layer}. The connectivity of the model corresponds to the one of a cortical slab under a surface area of 1 mm$^2$.

\subsection{NetPyNE implementation of the PDCM model}

NetPyNE employs a declarative language to specify the network parameters. Informally, declarative languages allow the user to describe 'what' they want, in contrast to imperative languages which specify 'how' to get there. In NetPyNE, this means the user only needs to provide the biological parameters at the different modeled scales, but is not required to implement all the low-level details. We therefore extracted the model parameters from the original PDCM publication \citep{potjans2012cell} and from the NEST source code available at OSB \citep{PDCode}. More specifically, the NetPyNE model specification required defining the parameters of 8 cell populations, 8 populations of spike generators (NetStims) that served as background inputs, and 68 connectivity rules. Since NetPyNE models require spatial dimensions, even if not explicitly used, we embedded the model into a cylinder of \SI{1470}{\micro\metre} depth and \SI{300}{\micro\metre} diameter, and set the cortical depth range (layer boundaries) for each population based on macaque V1 data \citep{schmidt2018multi}. 
Connectivity rules included the pre- and post-synaptic population, a fixed divergence value, and a weight and delay that followed a parameterized normal distribution. The model parameters were specified programmatically using NetPyNE's high-level declarative language, and could later be explored interactively via command line or NetPyNE's graphical user interface (GUI).

To reproduce the PDCM model, a new NEURON LIF neuron model  was required since the built-in LIF models do not allow setting the membrane time constant higher than the synaptic decay time constant. This feature was required to reproduce the original PDCM LIF model. We therefore implemented a new LIF point process neuron model using the NMODL (.mod) language.

The initial membrane potential for each neuron was set randomly from a Gaussian distribution with a mean of 58 mV and a standard deviation of 10 mV, as in the original article. However, we did not consider the initial transient phase of the first 100 ms of network activity in our analysis and took into account only the stationary condition of the network.

As in the original article, we implemented three different conditions in terms of the external inputs to the network \citep{potjans2012cell}:

1- Poisson and balanced: inputs follow a Poisson distribution and the number of external inputs to each population is balanced to generate a network behavior similar to that observed in biology.

2- Direct current (DC) input and balanced: inputs are replaced with an equivalent DC injection, and are balanced as in case 1. 

3- Poisson and unbalanced: inputs follow a Poisson distribution but each population receives the same number of inputs (unbalanced) resulting in non-biological firing rates, including absence of layer 6 excitatory activity.

The source code for the NetPyNE model, including the network Python code and the LIF neuron NMODL (.mod) code, are publicly available from  GitHub
\\(https://github.com/ceciliaromaro/PD\_in\_NetPyNE).

\subsection{Network rescaling}\label{section:rescaling}

The rescaling method implemented in our model was developed based on previous theoretical work 
\citep{van2015scalability, vreeswijk1998chaotic} and on the rescaling implementation of the original NEST PDCM model. The rescaling implementation of the original model is available as source code from the Open Source Brain (OSB) platform \citep{PDCode}, but was not described or addressed in the original article \citep{potjans2012cell}. Our rescaling implementation was simplified and adapted in order to guarantee the conservation of the first and second order statistics of network activity for all possible rescalings while being easy to implement in NetPyNE/NEURON. It is dependent on a single scaling parameter in the interval [0, 1], which is used to resize the number of network neurons, connections and external inputs as well as the synaptic weights, while keeping the matrix of connection probabilities and the proportions of cells per population fixed.

Since the original model article and source code did not include the rescaling option, we followed the rescaling implementation available in the OSB PDCM model version \citep{PDCode}. However, since the implementation methods were not described, we had to “scavenge” the source code to obtain the necessary information to understand the rescaling options and theoretical foundations \citep{van2015scalability,vreeswijk1998chaotic}. Our approach consisted in running the simulation  with different network sizes, adding breakpoints if necessary, in order to characterize all the relevant functions and parameters used for the rescaling mechanism. Their implementation allowed for different ways to rescale the network, most of them resulting in an alteration of the network statistics. We constrained our rescaling implementation to allow only for the specific cases in which the first and second order statistics are preserved. Therefore, we developed a simplified and consolidated rescaling function with a single scaling factor performing the following operations (see also Table \ref{tab:Rescale}):

1. Decrease the number of neurons and external inputs per neuron –- by multiplying them by the scale factor –- while keeping the proportions of cells per population fixed;

2. Decrease the number of connections per population –- by multiplying them by the square of the scale factor –- while keeping the probabilities of connections between populations unchanged;

3. Increase the synaptic weights –- by dividing them by the square root of the scale factor;

4. Provide each cell with an additional DC input current with a value corresponding to the total input lost due to rescaling.

\begin{table}[]
\begin{tabular}{lcc}
\hline
                                                                                              & \textbf{\begin{tabular}[c]{@{}c@{}}Full-scale\\ Network\end{tabular}} & \textbf{\begin{tabular}[c]{@{}c@{}}Resized \\ Network\end{tabular}}                                            \\ \hline
\textbf{Number of neurons}                                                                    & $N$                                                                     & $kN$                                                                                                             \\ \hline
\textbf{\begin{tabular}[c]{@{}l@{}}Number of \\ external inputs per neuron\end{tabular}}       & $I$                                                                     & $kI$                                                                                                             \\ \hline
\textbf{Probability of connection}                                                            & $p $                                                                    & $p $                                                                                                             \\ \hline
\textbf{\begin{tabular}[c]{@{}l@{}}Total connections \\ between two populations\end{tabular}} & $C_{i, j}$                                                           & $k^2C_{i, j}$                                                                                \\ \hline
\textbf{Synaptic weight}                                                                      & $W$                                                                     & $\frac{W}{\sqrt{k}} $                                                                                        \\ \hline
\textbf{Inner input per neuron}                                                               & $pN_{j}wf_{j \text{mean}} $                                              & $\sqrt{k}pN_{j}wf_{j \text{mean}}$                                                                                \\ \hline
\textbf{External input per neuron}                                                            & $IWf_{\text{external}}$                                                     & $\sqrt{k}IWf_{\text{external}}$                                                                                      \\ \hline
\textbf{DC input equivalence}                                                                 & $X$                                                                     & \begin{tabular}[c]{@{}c@{}}$X+(1-\sqrt{k})pN_{j}Wf_{j \text{mean}}$ \\ $+(1-\sqrt{k})IWf_{\text{external}}$\end{tabular} \\ \hline
\end{tabular}

\caption{Rescaling description. Transformation of the parameters of the full-scale network of size $N$ to a rescaled network of size $kN$.  $k$ is the scale factor, $f_{j-\text{mean}}$ is the mean firing rate of the pre-synaptic population and $f_{\text{external}}$ is the mean fire rate of the external population.}
\label{tab:Rescale}
\end{table}

The first three steps maintain the original network  proportions across layers, whereas the fourth step maintains the original statistics of network activity across layers. Therefore, this method is able to produce the same layer-specific average firing rates, synchrony and irregularity features in networks of smaller or larger size.

It is important to point out that to accurately reproduce the layer-specific average firing rates of the original model, it is fundamental to calculate the exact number of synapses and avoid using approximations \citep{REarticle}.

To compare the raster plot of spiking activity across the scaled networks we plotted approximately the same number of neurons as in the original publication, even though the total number of simulated neurons differed. Since the estimation of irregularity and synchrony may depend on the number of neurons included, we decided to always perform these calculations on a sample with the same number of neurons, despite comparing networks of different sizes. We utilized the same bin width (3 ms) as in the original article. The influence of the number of neurons is further assessed in the Results and Discussion sections.

Each population irregularity is estimated using a irregularity metric defined as the coefficient of variation (the estimated standard deviation divided by the mean) of the interspike interval (CV ISI), that is $$CV= \frac{\frac{1}{N}\sqrt{\sum_{i = 1}^{N}(x_i - \Bar{x})^2}}{\frac{1}{N}\sum_{i = 1}^N x_i},$$ where the sequence $x_i$ are the time intervals between the consecutive spikes of a fixed neuron (ISI). Synchrony per population is estimated as the variance of the spike count histogram normalized by its mean.

All the new model results and analyses were obtained using the NetPyNE tool, except for the synchrony statistic, which was calculated using the equation described in the original paper \citep{potjans2012cell}.

\section{Results}
\subsection{Reproduction of Potjans-Diesmann (PDCM) model results}

We start showing results from the NetPyNE implementation of the full scalle PDCM model with different sampling sizes and different external input conditions. The purpose is to compare the NetPyNe and NEST implementations and show that they are similar.

\begin{figure}[!ht]
\centering
{\includegraphics[width=0.6\textwidth]{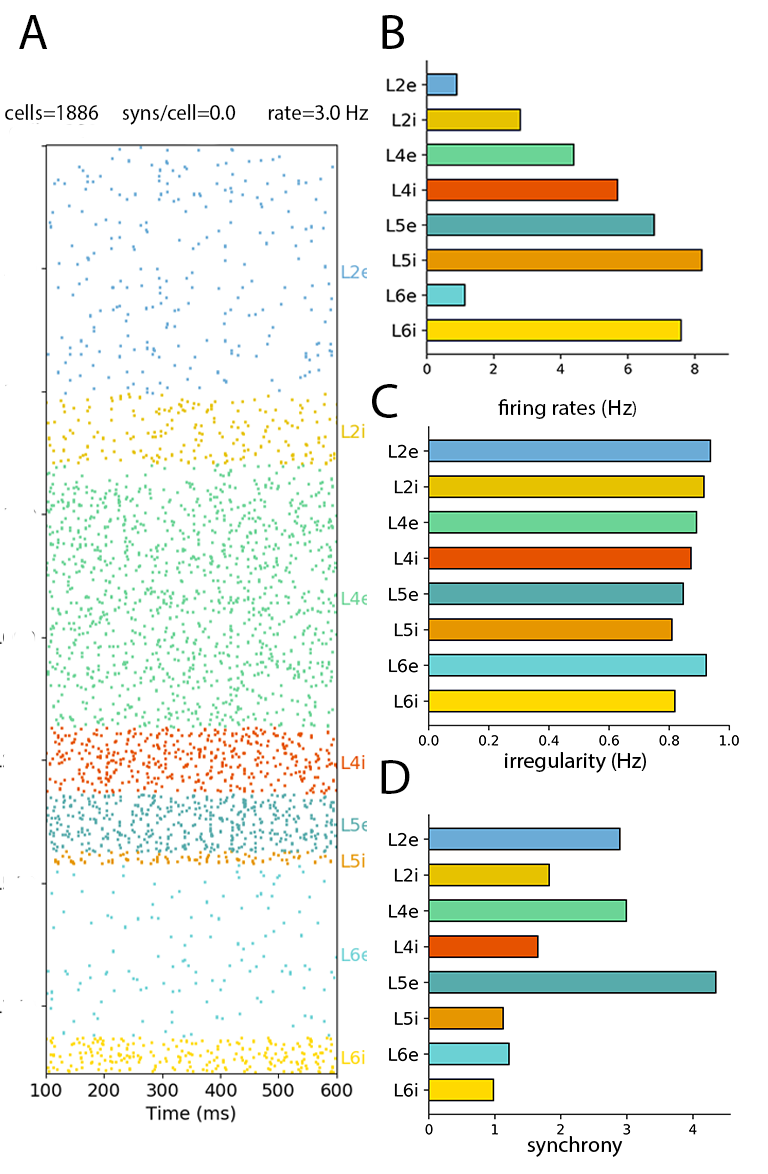}}
\caption{NetPyNE implementation results reproducing  Figure 6 of the original article \citep{potjans2012cell} (balanced Poisson inputs): (A) Raster plot of the 8 neural populations with 1862 excitatory and inhibitory neurons distributed across layers 2, 4, 5 and 6 for 500 ms. Only 2.3\% of the neurons in each populations are shown. (B) Mean firing rates of each cell population over 60 sec. (C) Irregularity per population estimated as the coefficient of variation of the interspike interval (CV ISI) over 60 sec. (D) Synchrony per population estimated as the variance of the spike count histogram normalized by its mean over 5 sec. Statistics in B, C and D were based on calculations with a fixed sample of 8000 neurons as explained in the Methods section.} 

\label{fig:PD_Original_Poisson}
\end{figure}

Figure \ref{fig:PD_Original_Poisson} shows results from the NetPyNE implementation reproducing the raster plot and firing rate, irregularity and synchrony statistics for the balanced Poisson inputs condition (Figure 6 of the original article \citep{potjans2012cell}). Although the results are not identical, due to the random components of the model (see Discussion), the major characteristics of the original model were reproduced: the raster plot included 1862 neurons and showed apparent asynchronous activity (but see below);  L2e and L6e exhibited the lowest firing rates with a mean around 1 Hz; L4e fired around 4 Hz, and L5e presented the higher excitatory firing rate, around 7 Hz. As in the original article, the irregularity of all populations was around 0.8, with the lowest irregularity (just under 0.8) was for L5i and L6i. The synchrony measure also closely matched the pattern across populations exhibited in the original model, with L5e showing the highest value, followed by L2e and L4e, and L5i and L6i displaying the lowest values.

A comparison of the mean population firing rates of the NetPyNE implementation with the original implementation (taken from Table 6 of the original article, which shows data only for the excitatory populations) is given in Table \ref{tab:HzNestPD}. The mean rates of the excitatory neurons of the NetPyNE implementation fall within the standard deviation ranges of their respective counterparts in the NEST implementation.

\begin{table}[h!]
\centering
\begin{adjustbox}{width=1\textwidth}
\begin{tabular}{lllllllll}
\hline
\textbf{Platform} & L2e  & L2i  & L4e  & L4i  & L5e  & L5i  & L6e  & L6i \\ \hline
\textbf{NEST}     & 0.85 &  -   & 4.45 &  -   & 7.59 &  -   & 1.09 &  - \\ \hline
\textbf{NetPyNE}  & 0.90 & 2.80 & 4.39 & 5.70 & 6.79 & 8.21 & 1.14 & 7.60
   \\ \hline
\textbf{NEST-100 trials}    & 1.11 $\pm$ 0.8 &  -   & 4.8 $\pm$ 1.1 &  -   & 11 $\pm$ 6.1 &  -   & 0.56 $\pm$ 0.9 &  -  \\ \hline   
\end{tabular}
\end{adjustbox}
\caption{Layer-specific firing rates (Hz) for the full version of the PDCM model implemented in NEST and NetPyNE. The third row shows mean and standard deviation firing rates from 100 runs of the NEST implementation with random numbers of external inputs to each layer (see \citep{potjans2012cell} for details).} \label{tab:HzNestPD}
\end{table}

\begin{figure}[h!]
\centering
{\includegraphics[width=0.6\textwidth]{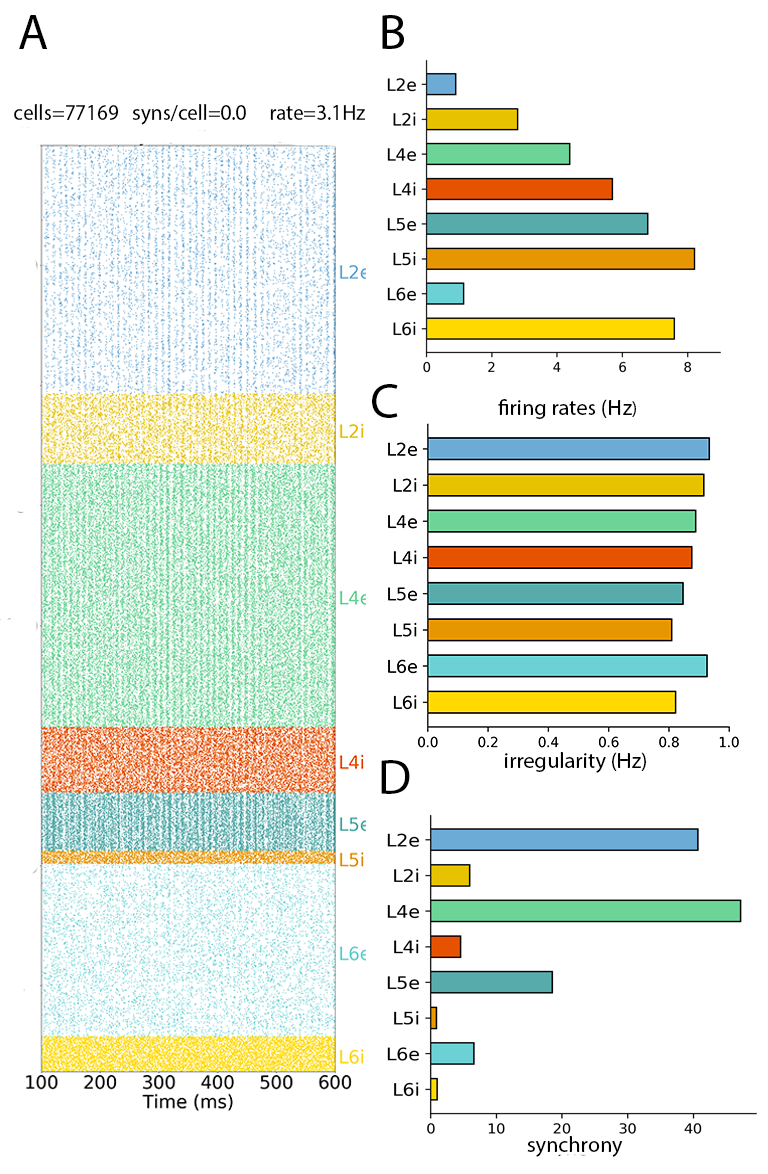}}
\caption{NetPyNE implementation of the PDCM model with full size sampling (balanced Poisson inputs). (A) Spike raster plot of the approximately 80k neurons distributed across layers 2, 4, 5 and 6 for 600ms. (B) Mean firing rates of each cell population over 60 sec. (C) Population irregularities, estimated as the coefficient of variation of the interspike interval, over a 60 sec simulation. (D) Synchrony per population estimated as the variance of the spike count histogram normalized by its mean over 5 sec. Statistics in B, C and D were based on calculations with the full number of neurons in the network.}
\label{fig:PD_Original_Poisson_Full}
\end{figure}

Next, we show that the ongoing spiking activity of the network and the corresponding synchrony measure depend on the number of neurons sampled. In Figure \ref{fig:PD_Original_Poisson_Full} we show results of the same full scale NetPyNE implementation of the PDCM model as in Figure \ref{fig:PD_Original_Poisson} but now with a sample of all neurons (77,169) in the network. The synchronous activity is visually obvious and the synchrony measure is strongly changed for all cell populations. On the other hand, the mean firing rate and the irregularity per cell population remained approximately the same. 

\begin{figure}[h!]
\centering
{\includegraphics[width=0.8\textwidth]{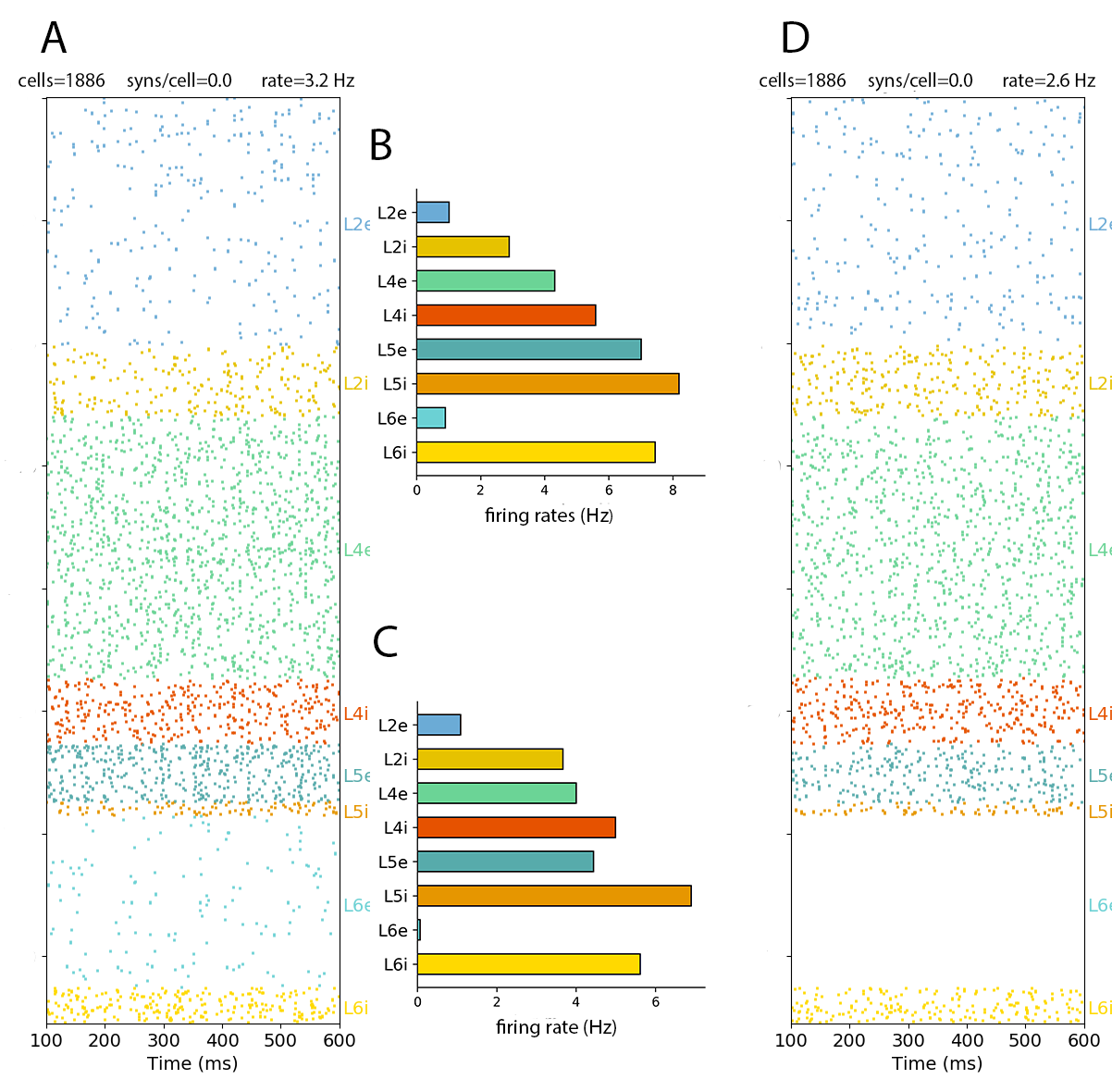}}
\caption{NetPyNE implementation results reproducing Figure 7 of the original article \citep{potjans2012cell}: (A) Raster plot of the 8 neural populations with 1862 excitatory and inhibitory neurons distributed across layers 2, 4, 5 and 6 for 500 ms with balanced DC inputs. Only 2.3\% of the neurons in each populations are shown. (B) Bar chart of single unit firing rates per population over 60 sec with balanced DC inputs. (C) Bar chart of single unit firing rates per population over 60 sec with unbalanced Poisson inputs. (D) Raster plot of the 8 neural populations with 1852 excitatory and inhibitory neurons distributed across layers 2, 4, 5 and 6 for 500 ms with unbalanced Poisson inputs. Only 2.3\% of the neurons in each populations are shown. Statistics in B, C and D were based on calculations with a fixed sample of 8000 neurons as explained in the Methods section.} 

\label{fig:PD_Original_DC}
\end{figure}

Figure \ref{fig:PD_Original_DC} reproduces the raster plot and mean firing rates for the DC current and unbalanced Poisson input conditions (panels A1, A2, B1 and B2 from Figure 7 in the original article \citep{potjans2012cell}). The raster plot was for a sample of 1862 neurons as in Figure \ref{fig:PD_Original_Poisson}. In similar fashion to the original article, replacing the balanced Poisson inputs with DC current did not affect the irregular firing displayed in the raster plot nor the population average firing rate properties. However, replacing them with unbalanced Poisson inputs resulted in no activity in L6e and modified the average firing rates across populations.

\clearpage

\subsection{Network rescaling}\label{NetworkRescaling}

Now that we have compared the full scale versions of the NetPyNE and NEST implementations with different sampling sizes, we will proceed to compare the rescaled NetPyNE implementations with the full scale NEST implementation.  

\begin{figure}[h!]
\centering
{\includegraphics[width=1.0\textwidth]{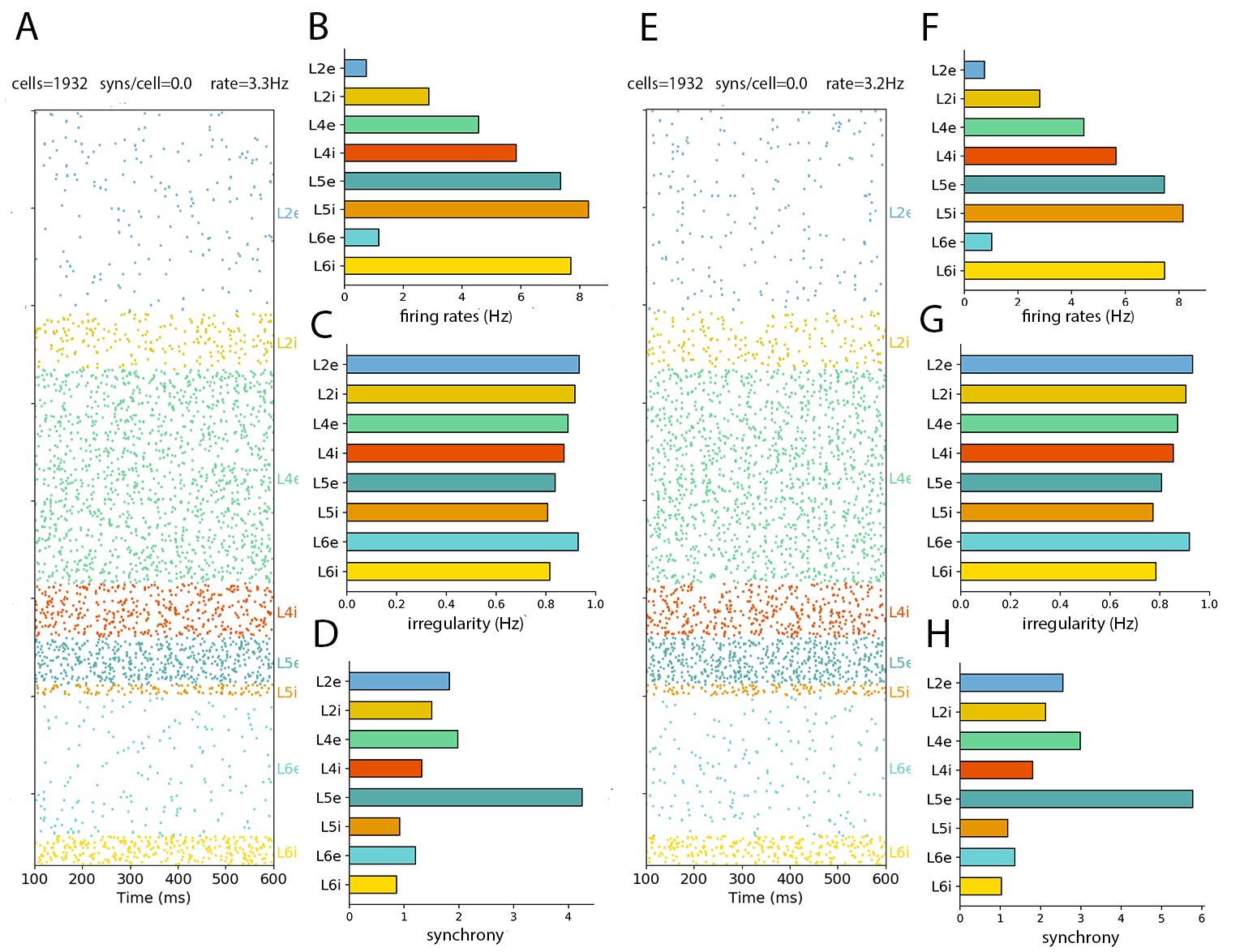}}
\caption{Network rescaled to 50\% of the number of total neurons. (A) Raster plot and (B--D) statistics for external Poisson input. (E) raster plot and (F--H) statistics for external DC current. The simulation times and number of neurons sampled and plotted were chosen as in Figure \ref{fig:PD_Original_Poisson}.}
\label{fig:50per}
\end{figure}

\begin{figure}[h!]
\centering
{\includegraphics[width=1.0\textwidth]{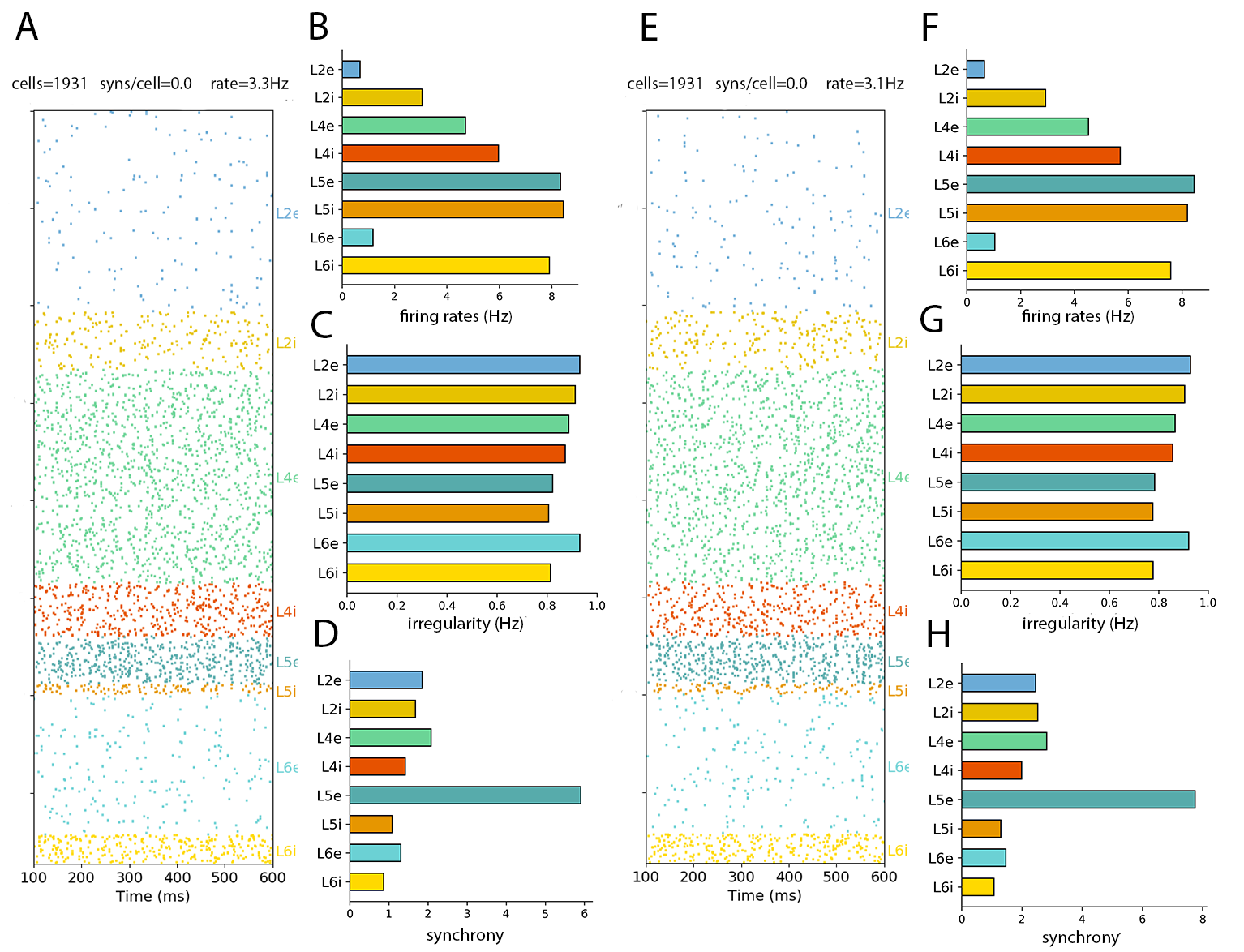}}
\caption{Network rescaled to 30\% of the number of total neurons. (A) Raster plot and (B--D) statistics for external Poisson input. (E) raster plot and (F--H) statistics for external DC current. The simulation times and number of neurons sampled and plotted were chosen as in Figure \ref{fig:PD_Original_Poisson}.}
\label{fig:30per}
\end{figure}

\begin{figure}[h!]
\centering
{\includegraphics[width=1.0\textwidth]{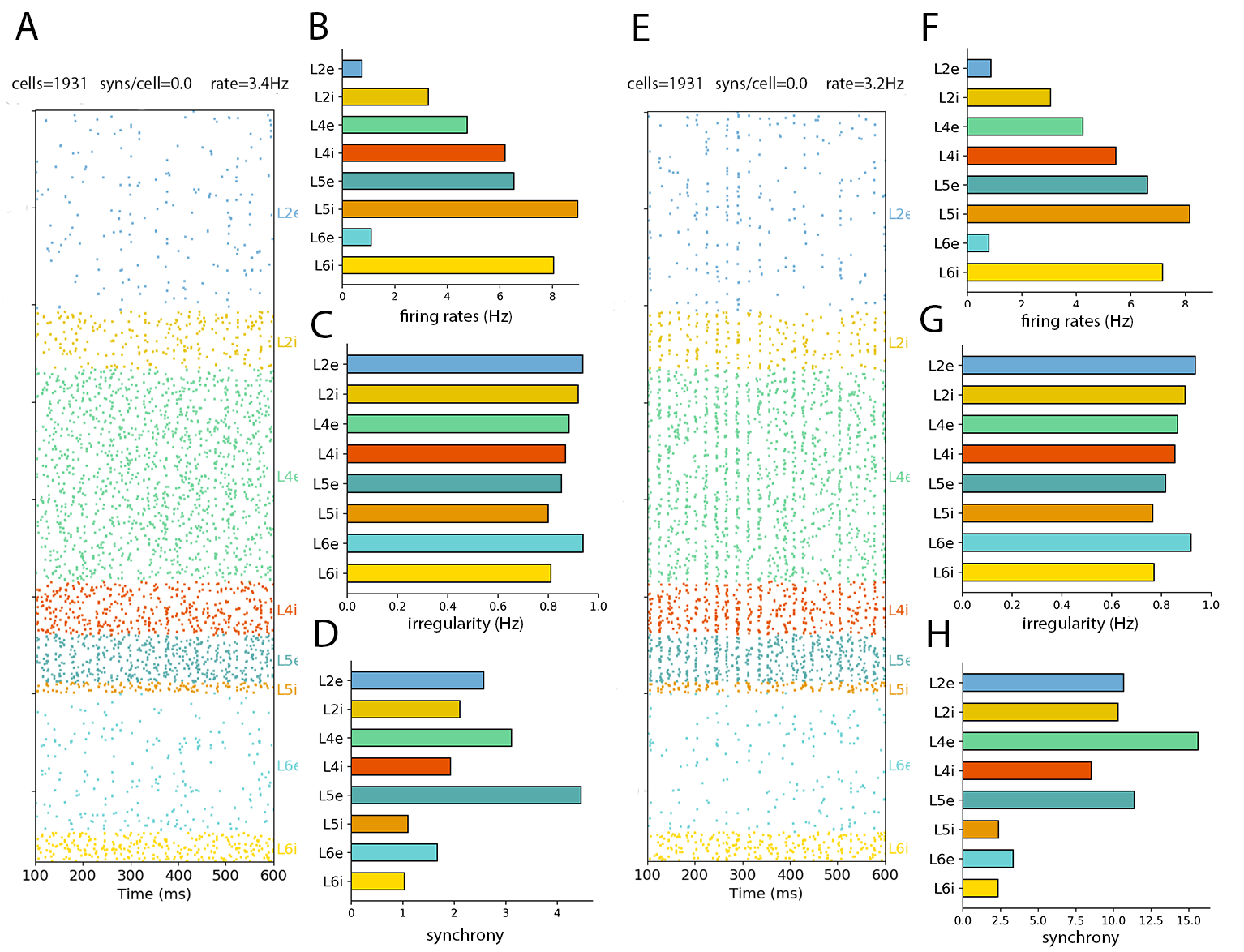}}
\caption{Network rescaled to 10\% of the number of total neurons. (A) Raster plot and (B--D) statistics for external Poisson input. (E) raster plot and (F--H) statistics for external DC current. The simulation times and number of neurons sampled and plotted were chosen as in Figure \ref{fig:PD_Original_Poisson}.}
\label{fig:10per}
\end{figure}

Figures \ref{fig:50per}-\ref{fig:10per} show raster plots and statistics for scaled down NetPyNE versions of the original PDCM network, with either Poisson or DC external inputs. As in the original article \citep{potjans2012cell}, raster plots show 1862 cells and the statistical measures were calculated using a fixed number of 8000 neurons. The raster plots exhibit similar firing patterns, and mean firing rate, irregularity and synchrony per layer as the full scale raster plot (Figure \ref{fig:PD_Original_Poisson}), both when using Poisson inputs (panels A-D of Figures \ref{fig:50per}, \ref{fig:30per} and \ref{fig:10per}) and DC inputs (panels E-H of Figures \ref{fig:50per} and \ref{fig:30per}). The raster plot and synchrony for the case of 10\% rescaling with DC external inputs differed from the results in Figure \ref{fig:PD_Original_Poisson}, as they exhibited a visually perceptible synchrony (Figure~\ref{fig:10per}E), and the synchrony values measured (Figure~\ref{fig:10per}H) were considerably higher.

\begin{figure}[h!]
\centering
{\includegraphics[width=1.0\textwidth]{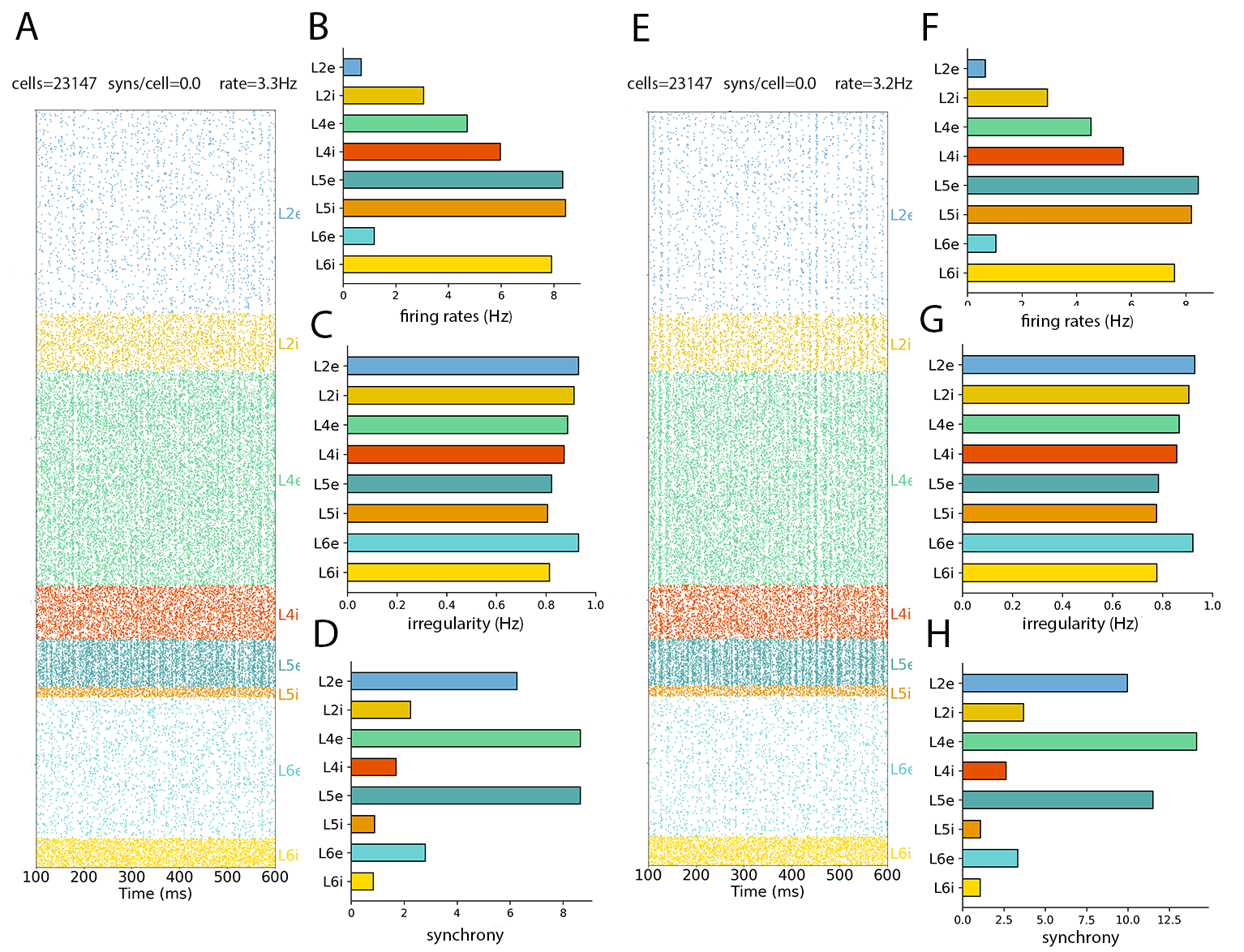}}
\caption{Network rescaled to 30\% of the total number of neurons with full size sampling. (A) Raster plot and (B--D) statistics for external Poisson input. (E) Raster plot and (F--H) statistics for external DC current. Raster plots and statistics based on all the neurons in the network.}
\label{fig:raster30_Full}
\end{figure}

To show again that the sampling size matters, Figure \ref{fig:raster30_Full} shows the results for the 30$\%$ rescaled network but includes all the neurons (23,147) in the raster plot and statistics calculations (compare to Figure \ref{fig:30per}). In a similar fashion to what was seen in the full scale network simulation (Figure~\ref{fig:PD_Original_Poisson_Full}), spike synchrony can be observed visually in the raster plots and the population synchrony values are significantly altered.

Extended results for the behavior of the cell populations mean firing rate, irregularity and synchrony as a function of the degree of rescaling and external input type for the NetPyNE implementation of the PDCM model are shown in Figure~\ref{fig:Statistics_rescaled}. A more complete set of data is given in Supplementary Tables \ref{tab:firingpoisson} (mean firing rates, Poisson input), \ref{tab:firingDC_current} (mean firing rate, DC input), \ref{tab:irregpoisson} (irregularity, Poisson input), \ref{tab:irregDC_current} (irregularity, DC input), \ref{tab:synpoisson2} (synchrony, Poisson input), and \ref{tab:synDC} (synchrony, DC input). They allow a comparison of the different rescaled NetPyNE implementations of the PDCM model with the original NEST implementation. They also allow a comparison of the NetPyNE implementations among themselves.

For both Poisson and DC external inputs, the mean population firing rates of all rescaled versions are close to the original results \citep{potjans2012cell}. For Poisson inputs, even extreme downscaling to 1\% resulted in mean firing rates within the ranges defined by the standard deviations calculated in the original article after 100 simulation trials (Table \ref{tab:firingpoisson}). For DC inputs, downscaling the network below 10\% resulted in no firing activity due to insufficient spiking input (low standard deviation of the local mean-field potential). Nevertheless, the mean rates of the DC input models with downscaling above 10\% also fall within the standard deviation intervals in the original article (Table \ref{tab:irregpoisson}).  

A comparison of the mean population firing rates over the NetPyNE models with different degrees of rescaling presents a relatively restrained variability (Figure~\ref{fig:Statistics_rescaled} and Tables \ref{tab:firingpoisson} and \ref{tab:firingDC_current}). The populations with larger firing variability are L2e and L2i but even for them the relative deviations in comparison to the full-scale network are generally below 30\%. In a comparison of input types, networks with Poisson inputs tend to exhibit larger mean firing variabilities at downscaling degrees below 10\%, while networks with DC inputs display large variabilities already at downscalings of 40\%.

The variability of the irregularity measure across the different levels of rescaling is much smaller than that of the mean firing rate, with relative deviations in relation to full scale around or below 1\% (Figure~\ref{fig:Statistics_rescaled} and Tables \ref{tab:irregpoisson} and \ref{tab:irregDC_current}). The populations with larger irregularity variability in comparison to the full scale network are L5e and L5i.

Finally, we also compared the network synchrony after rescaling. For this, we first compared three different sampling approaches to illustrate the effect of sample size in the calculation:

1. To sample a fixed percentage of neurons per population, totaling 8,000 neurons (the fixed percentage is simply given by the ratio between 8,000 and 77,169, which is the number of neurons in the full size network);

2. To sample 1,000 neurons per population, as in the original article, totaling 8,000 neurons;

3. To sample 2,000 neurons per population, totaling 16,000 neurons.

Differently from irregularity, synchrony depends on the sampling strategy adopted (see Table \ref{tab:synpoisson}). The observed discrepancies in synchrony may be a consequence of sampling a different number of neurons or a different percentage of the population size (see the Discussion section). Interestingly, with the exception of L5i and L6i, synchrony appears to increase linearly with
the number of sampled neurons from each population (Table \ref{tab:synpoisson}). For comparison, we show in Figure \ref{fig:Statistics_rescaled} the synchrony of each population for the full-scale and scaled down NetPyNE implementations using the sampling strategy of the original article (1,000 neurons per layer) (more details are given in Tables \ref{tab:synpoisson2} and \ref{tab:synDC}).

\begin{figure}[h!]
\center
{\includegraphics[width=\textwidth]{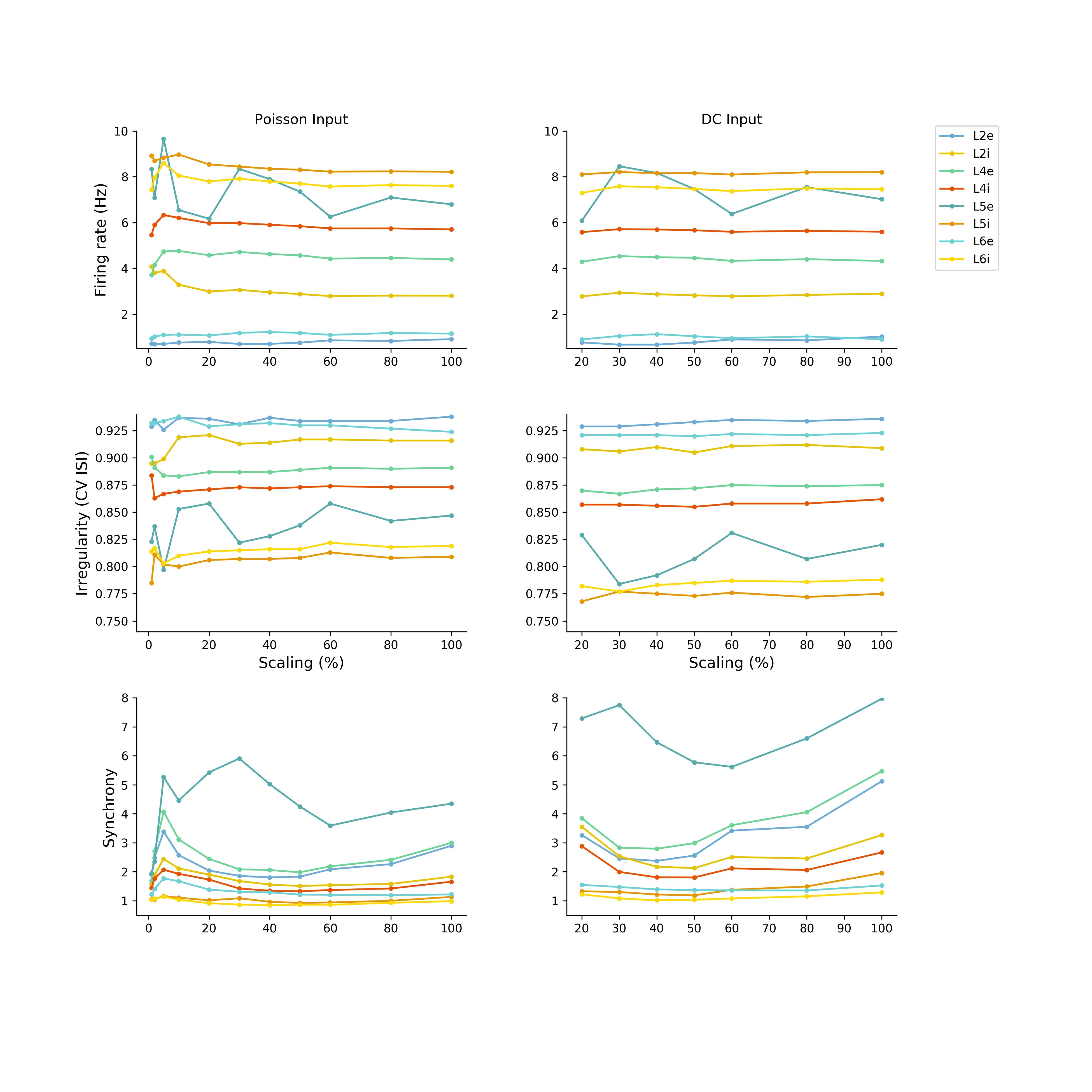}}
\caption{Mean population firing rate (top), irregularity (middle) and synchrony (bottom) of the 8 cell populations as a function of scaling for the NetPyNE reimplementation of the PDCM model, with Poisson (left) vs DC inputs (right). All results were calculated from 60-sec simulations and approximately 1000 neurons.\label{fig:Statistics_rescaled}}
\end{figure}

\begin{table}[h!]
\center{\begin{tabular}{lllllllll}
\hline
Population  & L2e & L2i & L4e & L4i& L5e & L5i& L6e & L6i 
\\ \hline Number of neurons  & 2,144  &   605  & 2,272  &  568  &  503  &  110  & 1492  &  306
\\ Synchrony  &   5.1  & 1.5  & 5.7  & 1.4  & 2.5  & 1.2  & 1.4  & 1.0
\\ \hline Number of neurons  & 1,000  &  1,000  & 1,000  & 1,000  & 1,000  & 1,000  & 1,000  & 1,000
\\ Synchrony  &   2.9  & 1.8  & 3.0  & 1.7  & 4.3  & 1.1  & 1.2  & 1.0
\\ \hline Number of neurons & 2,000  &  2,000  & 2,000  & 2,000  & 2,000  & 2,000  & 2,000  & 2,000
\\ Synchrony  &   4.9  & 2.7  & 5.1  & 2.3  & 7.9  & 1.1  & 1.6  & 1.0
\\ \hline
\end{tabular}
\caption{Synchrony of multi-unit spike trains quantified by the normalized variance of spike count histogram with bin width of 3 ms, based on different numbers of neurons and recorded during 60 s simulations. In the first row, the percentage of sampled neurons per population is fixed. In the second row, the number of sampled neurons per population is fixed. In both cases the total number of sampled neurons is 8,000. The third row is similar to the second with the number of sampled neurons per population doubled, totaling 16,000 neurons. For each row (separated from the others by continuous lines), the corresponding synchrony measures are given in the second line.\label{tab:synpoisson}}}
\end{table}

\subsection{Additional model analysis facilitated by NetPyNE}

Converting the PDCM model to the NetPyNE standardized specifications has the added advantage of allowing the user to readily make use of the tool's many built-in analysis functions. These range from 2D visualization of the cell locations to different representations of network connectivity to spiking activity and information flow measures. Importantly, these are available to the user through simple high-level function calls, which can be customized to include a specific time range, frequency range, set of populations, and visualization options.    

\begin{figure}[h!]
\centering
{\includegraphics[width=1\textwidth]{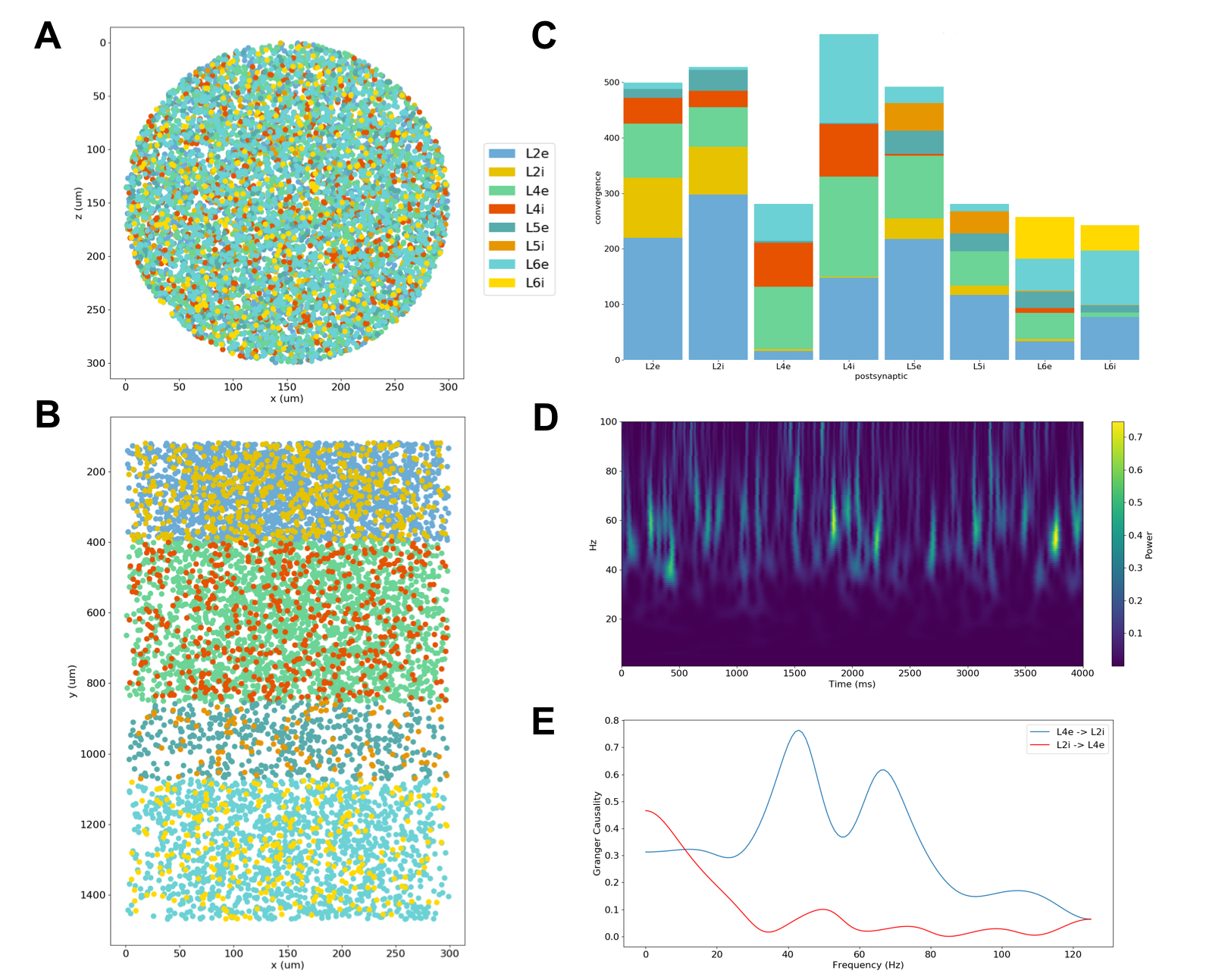}}
\caption{Analysis of model structure and simulation results facilitated by NetPyNE. (A) Top-down view (x-z plane) of network cell locations illustrates the diameter of the cylindrical volume modeled. (B) Side view (x-y plane) of network cell locations reveals cortical layer boundaries and populations per layer. Color indicates cell populations (see legend). (C) Stacked bar graph of the average connection convergence (number of presynaptic cells targeting each postsynaptic cell) for each population. (D) Spectrogram of average firing rate across all cells illustrating time-varying peaks in the gamma oscillation range. (E) Spectral Granger causality between L4e and L2i populations indicates stronger information flow from L4e to L2i than vice-versa.}
\label{fig:netpyne_figs}
\end{figure}

We illustrate the range of NetPyNE's analysis capabilities using the PDCM model reimplementation in NetPyNE \ref{fig:netpyne_figs}. All analyses were performed on the $10\%$-scaled version with 7713 cells and over 30M synapses, simulated for 4 biological seconds. First, we visualized the network cell locations from the top-down (Figure \ref{fig:netpyne_figs}A and side (Figure \ref{fig:netpyne_figs}B) views, which provided an intuitive representation of the cylindrical volume modeled, and the layer boundaries for each population. Next we plotted a stacked bar graph of convergence (Figure \ref{fig:netpyne_figs}C), a measure of connectivity that provides at-a-glance information on the average number and distribution of presynaptic inputs from each population. We then analyzed the spectrotemporal properties of the network's spiking activity through a Morlet wavelet-based spectrogram (Figure \ref{fig:netpyne_figs}D); results depicted time-varying broad frequency peaks in the gamma range (40-80 Hz), consistent with the largely irregular and asynchronous network activity. Finally, we measured the spectral Granger causality between L4e and L2e cells and found stronger information flow from L4e to L2i than vice-versa, particularly at gamma range frequencies, consistent with the canonical microcircuit \citep{Doug89}. Information flow analysis can reveal functional circuit pathways, including those involving inhibitory influences, that are not always reflected in the anatomical connectivity.

\section{Discussion}\label{sec:Discussion}

We reimplemented the PDCM model using the NetPyNE tool, a high-level interface to the NEURON simulator. The new model version reproduced the overall network dynamics of the original PDCM model, evaluated through population-specific average firing rates, irregularity and synchrony measures. The NetPyNE version also allows rescaling of the network size, while preserving the network statistics for most conditions. This feature can be used to study the effect of rescaling on network dynamics. For example, under certain conditions, network synchrony increased for smaller networks (see discussion below). Furthermore, the NetPyNE implementation (available in GitHub [link]) provides a clear separation of model parameters and implementation, facilitates extension of the model, for example to include more biophysically-realistic multicompartment neuron models, and enables employing NetPyNE’s analysis capabilities to gain further insights into the model. The latter was illustrated by visualizing the network's topology and connectivity, plotting the average firing rate spectrogram and calculating the spectral Granger causality (a measure of information flow) between two model populations (Figure \ref{fig:netpyne_figs}).

\subsection{Reproduction of original results}
We were able to reproduce all the network statistics (mean firing rate, irregularity and synchrony) for the three types of external inputs: balanced Poisson, DC current, and unbalanced Poisson -- compare Figure \ref{fig:PD_Original_Poisson} with Figure 6 of the original article \citep{potjans2012cell}), and Figure \ref{fig:PD_Original_DC} with Figure 7 of the original article. Notably, in the unbalanced Poisson input condition, we can observe both the ceased activity in L6e and the rate frequency changes in the other populations. Therefore, the NetPyNE PDCM model is able to effectively reproduce the model in the original article without loss or changes in the statistics.

\subsection{Preserved statistics in rescaled networks}

Our rescaling method works by keeping the random inputs unchanged on average \citep{van2015scalability} and fixing the proportion between the firing threshold and the square root of the number of connections \citep{vreeswijk1998chaotic} (see parameters in Table \ref{tab:Rescale}). This method managed to approximately preserve the mean firing rate and irregularity for all populations across all scaling percentages, ranging from 90\% to 1\% (Figure \ref{fig:Statistics_rescaled} and Tables \ref{tab:firingpoisson} to \ref{tab:irregDC_current}). The synchrony measure was similarly preserved for the Poisson external input condition (Table \ref{tab:synpoisson2}), but not for the DC input condition, as discussed below.

\subsection{Multiple factors affect synchrony}

The synchrony measure was dependent on the number of neurons used in its calculation, in general with a higher number of neurons resulting in higher synchrony values. Because of the different sampling strategies, comparing synchrony across populations and network models should be done with caution. For example, when we compared synchrony across populations sampling a fixed percentage of neurons per population (Table \ref{tab:synpoisson} top row) the two largest populations, namely L2e and L4e, exhibited the highest synchrony values. On the other hand, when we sampled a fixed number of neurons from each population, 1,000 as in the original article (Table \ref{tab:synpoisson} middle row) or 2,000 (Table \ref{tab:synpoisson} bottom row), the highest synchrony was displayed by population L5e. The strategy of sampling a fixed number of neurons per population also may lead to scaling distortions because a given fixed number corresponds to different percentages of the cell populations at each scaling degree. For example, in the full scale version 1,000 corresponds to almost 100\% of L5e neurons but to less than 5\% of L2e neurons. Besides, when the degree of downscaling is too strong ($\leq 20\%$) there may be not enough number of neurons in a population to include in the calculation.

In general, synchrony tends to decrease with the degree of downscale from the full size network (see plots at the bottom of Figure \ref{fig:Statistics_rescaled}). This is due to the increase in the DC current that we provide to neurons to compensate for the decrease in the number of connections (see Section \ref{section:rescaling}). This effect occurs up to a downscale level that depends on the population and external input type (between 40\% and 60\%). As we proceeded with the downscale past this point, we reached a situation in which the number of neurons was not sufficient to allow a reliable calculation of synchrony. For example, when we downscaled the networks to 10\% of the original size, we had to replace 99\% of the connections with DC inputs, and this resulted in large increases in synchrony (Figure \ref{fig:Statistics_rescaled}, bottom plots).

Another characteristic of synchrony is that it depends on the number of neurons sampled to do the statistics. For example, the raster plot and synchrony for the Poisson-driven full-scale network indicate, both visually and numerically, a high degree of synchrony when all neurons ($\sim$80k) are sampled (Figure \ref{fig:PD_Original_Poisson_Full}) but a very low degree of synchrony when only 2.3\% of the neurons are sampled (Figure \ref{fig:PD_Original_Poisson}). The same phenomenon was observed in the rescaled implementations. For example, the 30\% rescaled network displayed high synchrony when all neurons ($\sim$23k) were sampled (Figure \ref{fig:raster30_Full}) and low synchrony when a smaller subset of neurons ($\sim$2k) was sampled (Figure \ref{fig:30per}).

Synchrony was also dependent on the population average firing rate. The synchrony measure used (see Methods) increases with the heterogeneity of firing within the cell population \citep{pinsky1995synchrony}, which for equal population sizes and fixed bin size is higher for cells with higher firing rates. This dependence may be a possible explanation for the high synchrony of L5e neurons (Figure \ref{fig:Statistics_rescaled} bottom plots; see also Tables \ref{tab:synpoisson2} and \ref{tab:synDC}).

Synchrony was generally higher under the DC input condition than the Poisson input condition (Figure \ref{fig:Statistics_rescaled} bottom plots; see also Tables \ref{tab:synpoisson2} and \ref{tab:synDC}). We hypothesize this is due to the two sources of randomization present in the Poisson-driven network: the Poisson inputs and the random pattern of connection. In the DC condition we removed the Poisson inputs, thus increasing the network synchrony. For very high downscaling, e.g. 10\%, synchrony becomes visually perceptible in the raster plot for DC inputs but not in the one for Poisson inputs (compare Figures \ref{fig:10per}H and \ref{fig:10per}D). This effect is not seen for intermediate downscaling levels, cf. Figures \ref{fig:50per}D,H (50\% downscaling) and \ref{fig:30per}D,H (30\% downscaling) because the fraction of sampled cells is not high enough.

The rescaling method used here has the theoretical property of not adding synchrony or regularity to asynchronous irregular networks  \citep{vreeswijk1998chaotic, van2015scalability}. In our study, we found that irregularity and synchrony did not appear to be affected by rescaling up to the limits for which mean firing rates were within the standard deviations of the original article \citep{potjans2012cell}, namely  1\% for Poisson inputs and 10\% for DC inputs.

\section*{Acknowledgments}
This work was produced as part of the activities of NIH U24EB028998, R01EB022903, U01EB017695, NYS SCIRB DOH01-C32250GG-3450000, NSF 1904444; and FAPESP (S. Paulo Research Foundation) Research, Disseminations and Innovation Center for Neuromathematics 2013/07699-0, 2015/50122-0, 2018/20277-0. C.R. (grant number 88882.378774/2019-01) and F.A.N/ (grant number 88882.377124/2019-01) are recipients of PhD scholarships from the Brazilian Coordenação de Aperfeiçoamento de Pessoal de Nível Superior (CAPES). A.C.R. is partially supported by a CNPq
fellowship (grant 306251/2014-0).

\pagebreak

\beginsupplement

\section{Supplementary Material}

\begin{table}[h!]
\centering
\begin{adjustbox}{width=1\textwidth}
\begin{tabular}{lllllllll}
\hline
Population  & L2e & L2i & L4e & L4i & L5e & L5i & L6e & L6i
\\ Scaling \\ \hline
100\%	&	0.90		&	2.80		&	4.39		&	5.70		&	6.80	&	8.22	&	1.14				&	1.14
\\80\%	&	0.82 (9\%) & 2.81 (0\%) & 4.45 (1\%) & 5.74 (1\%) & 7.10 (4\%) & 8.24 (0\%) & 1.17 (3\%) & 1.17 (3\%)
\\60\%	&	0.85 (6\%) & 2.79 (0\%) & 4.42 (1\%) & 5.74 (1\%) & 6.25 (8\%) & 8.22 (0\%) & 1.09 (4\%) & 1.09 (4\%)
\\	\hline														
50\%	&	0.75 (17\%) & 2.87 (3\%) & 4.57 (4\%) & 5.84 (2\%) & 7.35 (8\%) & 8.31 (1\%) & 1.17 (3\%) & 1.17 (3\%)
\\40\%	&	0.70 (22\%) & 2.95 (5\%) & 4.62 (5\%) & 5.90 (4\%) & 7.90 (16\%) & 8.35 (2\%) & 1.22 (7\%) & 1.22 (7\%)
\\30\%	&	\textbf{0.69 (23\%)} & 3.06 (9\%) & 4.71 (7\%) & 5.98 (5\%) & 8.34 (23\%) & 8.45 (3\%) & 1.18 (4\%) & 1.18 (4\%)
\\20\%	&	0.78 (13\%) & 2.99 (7\%) & 4.57 (4\%) & 5.97 (5\%) & 6.17 (9\%) & 8.54 (4\%) & 1.06 (7\%) & 1.06 (7\%)
\\10\%	&	0.75 (17\%) & 3.28 (17\%) & 4.76 (8\%) & 6.20 (9\%) & 6.55 (4\%) & \textbf{8.97 (9\%)} & 1.10 (4\%) & 1.10 (4\%)
\\	\hline														
5\%	&	\textbf{0.69 (23\%)} & 3.88 (39\%) & 4.74 (8\%) & \textbf{6.33 (11\%)} & \textbf{9.66 (42\%)} & 8.83 (7\%) & 1.09 (4\%) & 1.09 (4\%)
\\2 \%	&	\textbf{0.69 (23\%)} & 3.8 (36\%) & 4.15 (5\%) & 5.90 (4\%) & 7.08 (4\%) & 8.70 (6\%) & 1.02 (11\%) & 1.02 (11\%)
\\1\%	&	0.71 (21\%) & \textbf{4.08 (46\%)} & \textbf{3.71 (15\%)} & 5.46 (4\%) & 8.33 (23\%) & 8.92 (9\%) & \textbf{0.93 (18\%)} & \textbf{0.93 (18\%)}
\\ \hline
NEST-100 trials    & 1.11 $\pm$ 0.8 &  -   & 4.8 $\pm$ 1.1 &  -   & 11 $\pm$ 6.1 &  -   & 0.56 $\pm$ 0.9 &  - \\ \hline
\end{tabular}
\end{adjustbox}
\caption{Population mean firing rates (Hz) for rescaled NetPyNE versions of the original PDCM model with Poisson external input. All results calculated from 60-sec simulations and approximately 1850 neurons. Relative deviations in relation to the full-scale NetPyNE version $\left( \left| f_{\mbox{x\%}} - f_{100\%} \right|/f_{100\%} \right)$, are shown within parentheses, and their maxima in bold. For comparison, the last row shows mean and standard deviation firing rates from 100 runs of the NEST implementation with random numbers of external inputs to each layer (see \citep{potjans2012cell} for details). \label{tab:firingpoisson}
}
\end{table}

\begin{table}[h!]
\centering
\begin{adjustbox}{width=1\textwidth}
\begin{tabular}{lllllllll}
\hline
Population  & L2e & L2i & L4e & L4i& L5e & L5i& L6e & L6i 
\\ Scaling \\ \hline
100\%	&	1.02		&	2.89	&	4.32	&	5.60	&	7.02	&	8.20	&	0.90				&	0.90			
\\80\%	&	0.85 (17\%) & 2.83 (2\%) & 4.40 (2\%) & 5.64 (1\%) & 7.55 (8\%) & 8.20 (0\%) & 1.02 (13\%) & 1.02 (13\%)
\\60\%	&	0.89 (13\%) & 2.77 (4\%) & 4.32 (0\%) & 5.59 (0\%) & 6.38 (9\%) & 8.10 (1\%) & 0.94 (4\%) & 0.94 (4\%)
\\	\hline													
50\%	&	0.75 (26\%) & 2.82 (2\%) & 4.46 (3\%) & 5.66 (1\%) & 7.46 (6\%) & 8.16 (0\%) & 1.03 (14\%) & 1.03 (14\%)
\\40\%	&	\textbf{0.66 (35\%)} & 2.86 (1\%) & 4.49 (4\%) & 5.70 (2\%) & 8.17 (16\%) & 8.16 (0\%) & \textbf{1.12 (24\%)} & \textbf{1.12 (24\%)}
\\30\%	&	\textbf{0.66 (35\%)} & 2.93 (1\%) & \textbf{4.53 (5\%)} & \textbf{5.71 (2\%)} & \textbf{8.46 (21\%)} & 8.21 (0\%) & 1.05 (17\%) & 1.05 (17\%)
\\20\%	&	0.76 (25\%) & 2.77 (4\%) & 4.29 (1\%) & 5.58 (0\%) & 6.08 (13\%) & \textbf{8.11 (1\%)} & 0.89 (1\%) & 0.89 (1\%)
\\10\%	&	0.87 (15\%) & \textbf{3.06 (6\%)} & 4.25 (2\%) & 5.46 (2\%) & 6.62 (6\%) & 8.17 (0\%) & 0.80 (11\%) & 0.80 (11\%)
\\ \hline
NEST-100 trials   & 1.11 $\pm$ 0.8 &  -   & 4.8 $\pm$ 1.1 &  -   & 11 $\pm$ 6.1 &  -   & 0.56 $\pm$ 0.9 &  - \\ \hline
\end{tabular}
\end{adjustbox}
\caption{Population firing rates (Hz) for rescaled NetPyNE versions of the original PDCM model with DC current input. All results calculated from 60-sec simulations and approximately 1850 neurons. Relative deviations in relation to the full-scale NetPyNE version $\left( \left| f_{\mbox{x\%}} - f_{100\%} \right|/f_{100\%} \right)$ are shown within parentheses, and their maxima in bold. For comparison, the last row shows mean and standard deviation firing rates from 100 runs of the NEST implementation with random numbers of external inputs to each layer (see \citep{potjans2012cell} for details).
\label{tab:firingDC_current}}
\end{table}

\pagebreak

\begin{table}[h!]
\centering
\begin{adjustbox}{width=1\textwidth}
\begin{tabular}{lllllllll}
\hline
Population  & L2e & L2i & L4e & L4i& L5e & L5i& L6e & L6i 
\\ Scaling \\ \hline
100\%   & 0.938   & 0.916   & 0.891   & 0.873   & 0.847   & 0.809   & 0.924   & 0.819 
\\80\%  & \textbf{0.934 (0.45\%)}  & 0.916 (0.06\%) & 0.890 (0.17\%)  & 0.873 (0.07\%) & 0.842 (0.63\%) & 0.808 (0.15\%)  & 0.927 (0.29\%) & 0.818 (0.19\%)
\\60\%  & \textbf{0.934 (0.45\%)}  & 0.917 (0.04\%)  & 0.891 (0.02\%)  & 0.874 (0.12\%)  & 0.858 (1.25\%)  & 0.813 (0.48\%) & 0.930 (0.63\%)  & 0.822 (0.27\%)
\\ \hline
  50\%  & \textbf{0.934 (0.39\%)}  & 0.917 (0.06\%)  & 0.889 (0.19\%)  & 0.873 (0.06\%)  & 0.838 (1.12\%)  & 0.808 (0.24\%)  & 0.930 (0.68\%)  & 0.816 (0.38\%)
\\40\%  & 0.937 (0.08\%)  & 0.914 (0.31\%)  & 0.887 (0.44\%)  & 0.872 (0.12\%)  & 0.828 (2.30\%) & 0.807 (0.36\%) & 0.932 (0.88\%) & 0.816 (0.40\%)
\\30\%  & \textbf{0.931 (0.72\%)}  & 0.913 (0.37\%) & 0.887 (0.45\%) & 0.873 (0.06\%) & 0.822 (2.95\%) & 0.807 (0.32\%) & 0.931 (0.72\%) & 0.815 (0.60\%)
\\20\%  & 0.936 (0.15\%)  & 0.921 (0.50\%)  & 0.887 (0.41\%) & 0.871 (0.28\%) & 0.858 (1.22\%) & 0.806 (0.41\%) & 0.929 (0.55\%) & 0.814 (0.64\%)
\\10\%  & 0.937 (0.05\%)  & 0.919 (0.31\%) & 0.883 (0.87\%) & 0.869 (0.51\%) & 0.853 (0.62\%) & 0.800 (1.23\%) & \textbf{0.938 (1.55\%)} & 0.810 (1.11\%)
\\ \hline
   5\%  & \textbf{0.926 (1.21\%)} & 0.899 (1.90\%) & 0.884 (0.82\%) & 0.867 (0.76\%) & \textbf{0.797 (5.99\%)} & 0.802 (0.97\%) & 0.934 (1.03\%) & \textbf{0.803 (1.97\%)}
\\ 2\%  & 0.935 (0.32\%) & \textbf{0.895 (2.34\%)} & 0.891 (0.03\%) & \textbf{0.863 (1.21\%)} & 0.837 (1.23\%) & 0.811 (0.15\%) & 0.932 (0.81\%) & 0.817 (0.24\%)
\\ 1\%  & \textbf{0.929 (0.92\%)} & 0.895 (2.30\%) & \textbf{0.901 (1.12\%)} & 0.884 (1.16\%) & 0.823 (2.90\%) & \textbf{0.785 (2.98\%)} & 0.932 (0.82\%) & 0.814 (0.66\%) 
\\ \hline
\end{tabular}
\end{adjustbox}
\caption{Irregularity of single-unit spike trains for rescaled NetPyNE versions of the original PDCM model with Poisson external input. All results calculated from 60-sec simulations and approximately 1000 neurons per population. Relative deviations in relation to the full-scale NetPyNE version $\left( \left| f_{\mbox{x\%}} - f_{100\%} \right|/f_{100\%} \right)$ are shown within parentheses, and their maxima in bold.\label{tab:irregpoisson}}
\end{table}

\begin{table}[h!]
\centering
\begin{adjustbox}{width=1\textwidth}
\begin{tabular}{lllllllll}
\hline
Population  & L2e & L2i & L4e & L4i& L5e & L5i& L6e & L6i 
\\ Scaling \\ \hline
100\%	&	0.936			&	0.909				&	0.875				&	0.862				&	0.820				&	0.775				&	0.923				&	0.788			
\\80\%	&	0.934 (0.21\%)	&	0.912 (0.33\%)	&	0.874	 (0.11\%)	&	0.858	 (0.46\%)	&	0.807	 (1.59\%)	&	0.772	 (0.39\%)	&	0.921	 (0.22\%)	&	0.786	 (0.25\%)
\\60\%	&	0.935 (0.11\%)	&	0.911 (0.22\%)	&	0.875	 (0.00\%)	&	0.858	 (0.46\%)	&	0.831	 (1.34\%)	&	0.776	 (0.13\%)	&	0.922	 (0.11\%)	&	0.787	 (0.13\%)
\\	\hline													
50\%	&	0.933 (0.32\%)	&	0.905 (0.44\%)	&	0.872	 (0.34\%)	&	\textbf{0.855	 (0.81\%)}	&	0.807	 (1.59\%)	&	0.773	 (0.26\%)	&	\textbf{0.920	 (0.33\%)}	&	0.785	 (0.38\%)
\\40\%	&	0.931 (0.53\%)	&	0.910 (0.11\%)	&	0.871	 (0.46\%)	&	0.856	 (0.70\%)	&	0.792	 (3.41\%)	&	0.775	 (0.00\%)	&	0.921	 (0.22\%)	&	0.783	 (0.63\%)
\\30\%	&	\textbf{0.929 (0.75\%)}	&	0.906 (0.33\%)	&	0.867	 (0.91\%)	&	0.857	 (0.58\%)	&	\textbf{0.784	 (4.39\%)}	&	0.777	 (0.26\%)	&	0.921	 (0.22\%)	&	0.777	 (1.40\%)
\\20\%	&	\textbf{0.929 (0.75\%)}	&	0.908 (0.11\%)	&	0.870	 (0.57\%)	&	0.857	 (0.58\%)	&	0.829	 (1.10\%)	&	0.768	 (0.90\%)	&	0.921	 (0.22\%)	&	0.782	 (0.76\%)
\\10\%	&	0.937 (0.11\%)	&	\textbf{0.896 (1.43\%)}	&	\textbf{0.866	 (1.03\%)}	&	\textbf{0.855	 (0.81\%)}	&	0.817	 (0.37\%)	&	\textbf{0.766	 (1.16\%)}	&	\textbf{0.920	 (0.33\%)}	&	\textbf{0.771	 (2.16\%)}
\\ \hline
\end{tabular}
\end{adjustbox}
\caption{Irregularity of single-unit spike trains for rescaled NetPyNE versions of the original PDCM model with DC current input. All results calculated from 60-sec simulations and approximately 1000 neurons per population. Relative deviations in relation to the full-scale NetPyNE version $\left( \left| f_{\mbox{x\%}} - f_{100\%} \right|/f_{100\%} \right)$ are shown within parentheses, and their maxima in bold.\label{tab:irregDC_current}}
\end{table}

\pagebreak

\begin{table}[h!]
\centering
\begin{adjustbox}{width=1\textwidth}
\begin{tabular}{lllllllll}
\hline
Layer  & L2e & L2i & L4e & L4i& L5e & L5i& L6e & L6i 
\\ Scaling \\ \hline
100\%	&	2.9	 	&	1.8			&	3.0		&	1.7		&	4.4		&	1.1		&	1.2		&	1.0			
\\80\%	&	2.3 (21\%) & 1.6 (11\%) & 2.4 (20\%) & 1.4 (18\%) & 4.0 (9\%) & 1.0 (9\%) & 1.2 (0\%) & 0.9 (10\%)
\\60\%	&	2.1 (28\%) & 1.5 (17\%) & 2.2 (27\%) & 1.4 (18\%) & 3.6 (18\%) & \textbf{0.9 (18\%)} & 1.2 (0\%) & 0.9 (10\%)
\\	\hline														
50\%	&	\textbf{1.8 (38\%)} & 1.5 (17\%) & 2.0 (33\%) & 1.3 (24\%) & 4.3 (2\%) & \textbf{0.9 (18\%)} & 1.2 (0\%) & 0.9 (10\%)
\\40\%	&	\textbf{1.8 (38\%)} & 1.6 (11\%) & 2.1 (30\%) & 1.4 (18\%) & 5.0 (14\%) & 1.0 (9\%) & 1.3 (8\%) & \textbf{0.8 (20\%)}
\\30\%	&	1.9 (34\%) & 1.7 (6\%) & 2.1 (30\%) & 1.4 (18\%) & 5.9 (34\%) & 1.1 (0\%) & 1.3 (8\%) & 0.9 (10\%)
\\20\%	&	2.0 (31\%) & 1.9 (6\%) & 2.4 (20\%) & 1.7 (0\%) & 5.4 (23\%) & 1.0 (9\%) & 1.4 (17\%) & 0.9 (10\%)
\\10\%	&	2.6 (10\%) & 2.1 (17\%) & 3.1 (3\%) & 1.9 (12\%) & 4.5 (2\%) & 1.1 (0\%) & 1.7 (42\%) & 1.0 (0\%)
\\	\hline
5\%	&	3.4 (17\%) & \textbf{2.4 (33\%)} & 4.1 (37\%) & \textbf{2.1 (24\%)} & 5.3 (20\%) & 1.2 (9\%) & \textbf{1.8 (50\%)} & 1.1 (10\%)
\\2 \%	&	2.5 (14\%) & 1.9 (6\%) & 2.7 (10\%) & 1.8 (6\%) & 2.4 (45\%) & 1.1 (0\%) & 1.4 (17\%) & 1.1 (10\%)
\\1\%	&	1.9 (34\%) & 1.6 (11\%) & \textbf{1.7 (43\%)} & 1.4 (18\%) & \textbf{2.0 (55\%)} & 1.1 (0\%) & 1.2 (0\%) & 1.0 (0\%)
\\ \hline
\end{tabular}
\end{adjustbox}
\caption{Synchrony of multi-unit spike trains for rescaled NetPyNE versions of the original PDCM model with Poisson external input. All results calculated from 60-sec simulations and approximately 1000 neurons per population. Relative deviations in relation to the full-scale NetPyNE version $\left( \left| f_{\mbox{x\%}} - f_{100\%} \right|/f_{100\%} \right)$ are shown within parentheses, and their maxima in bold.\label{tab:synpoisson2}}
\end{table}

\begin{table}[h!]
\centering
\begin{adjustbox}{width=1\textwidth}
\begin{tabular}{lllllllll}
\hline
Layer  & L2e & L2i & L4e & L4i& L5e & L5i& L6e & L6i 
\\ Scaling \\ \hline
100\%	&	5.1		&	3.3			&	5.5			&	2.7		&	8.0		&	2.0		&	1.5		&	1.3			
\\80\%	&	3.6 (29\%) & 2.5 (24\%) & 4.1 (25\%) & 2.1 (22\%) & 6.6 (18\%) & 1.5 (25\%) & 1.4 (7\%) & 1.2 (8	\%)
\\60\%	&	3.4 (33\%) & 2.5 (24\%) & 3.6 (35\%) & 2.1 (22\%) & 5.6 (30\%) & 1.4 (30\%) & 1.4 (7\%) & 1.1 (15	\%)
\\	\hline														
50\%	&	2.6 (49\%) & 2.1 (36\%) & 3.0 (45\%) & 1.8 (33\%) & 5.8 (28\%) & \textbf{1.2 (40\%)} & 1.4 (7\%) & 1.0 (23\%)
\\40\%	&	2.4 (53\%) & 2.2 (33\%) & 2.8 (49\%) & 1.8 (33\%) & 6.5 (19\%) & \textbf{1.2 (40\%)} & 1.4 (7\%) & 1.0 (23\%)
\\30\%	&	2.5 (51\%) & 2.5 (24\%) & 2.8 (49\%) & 2.0 (26\%) & 7.8 (3\%) & 1.3 (35\%) & 1.5 (0\%) & 1.1 (15\%)
\\20\%	&	3.3 (35\%) & 3.6 (9\%) & 3.9 (29\%) & 2.9 (7\%) & 7.3 (9\%) & 1.3 (35\%) & 1.6 (7\%) & 1.2 (8\%)
\\10\%	&	\textbf{10.7 (110\%)} & \textbf{10.3 (212\%)} & \textbf{15.6 (184\%)} & \textbf{8.5 (215\%)} & \textbf{11.4 (43\%)} & 2.4 (20\%) & \textbf{3.3 (120\%)} & \textbf{2.3 (77\%)}
\\ \hline
\end{tabular}
\end{adjustbox}
\caption{Synchrony of multi-unit spike trains for rescaled NetPyNE versions of the original PDCM model with DC current input. All results calculated from 60-sec simulations and approximately 1000 neurons per population. Relative deviations in relation to the full-scale NetPyNE version $\left( \left| f_{\mbox{x\%}} - f_{100\%} \right|/f_{100\%} \right)$ are shown within parentheses, and their maxima in bold.\label{tab:synDC}}
\end{table}

\end{document}